\documentstyle[11pt,epsf]{article}

\input psfig
\def\beq{\begin{eqnarray}}
\def\eeq{\end{eqnarray}}
\def\bea{\begin{eqnarray*}}
\def\eea{\end{eqnarray*}}

\def\centeron#1#2{{\setbox0=\hbox{#1}\setbox1=\hbox{#2}\ifdim
\wd1>\wd0\kern.5\wd1\kern-.5\wd0\fi
\copy0\kern-.5\wd0\kern-.5\wd1\copy1\ifdim\wd0>\wd1
\kern.5\wd0\kern-.5\wd1\fi}}
\def\ltap{\;\centeron{\raise.35ex\hbox{$<$}}{\lower.65ex\hbox{$\sim$}}\;}
\def\gtap{\;\centeron{\raise.35ex\hbox{$>$}}{\lower.65ex\hbox{$\sim$}}\;}

\def\singleandthirdspaced{\baselineskip=\normalbaselineskip\multiply
    \baselineskip by 130\divide\baselineskip by 100}
\def\singlespaced{\baselineskip=\normalbaselineskip}

\newcommand{\newc}{\newcommand}
\newc{\qbar}{{\overline q}}
\newc{\Kahler}{K\"ahler }
\newc{\deltaGS}{\delta_{\rm GS}}
\begin{document}
\begin{titlepage}
\begin{flushright}
{\large hep-th/9905219 \\ SCIPP-99/18\\

}
\end{flushright}

\vskip 1.2cm

\begin{center}

{\LARGE\bf Possible Scales of New Physics}\footnote{Plenary talk
on Beyond the Standard Model Physics, DPF99 (UCLA).}

\vskip 1.4cm

{\large Michael Dine}
\\
\vskip 0.4cm
%{\it $^a$Stanford Linear Accelerator Center,
%     Stanford CA 94309} \\
{\it Santa Cruz Institute for Particle Physics,
     Santa Cruz CA 95064  } \\
%{\it $^c$Physics Department,
%     University of California,
%     Santa Cruz CA 95064  } \\

\vskip 4pt

\vskip 1.5cm

\begin{abstract}
The biggest question in beyond the standard model physics is what
are the scales of new physics.   Ideas about scales, as well as
experimental evidence and constraints, are surveyed for a variety
of possible forms of new physics:  supersymmetry, neutrino masses,
unification, and superstring theory.
\end{abstract}

\end{center}

\vskip 1.0 cm

\end{titlepage}
\setcounter{footnote}{0} \setcounter{page}{2}
\setcounter{section}{0} \setcounter{subsection}{0}
\setcounter{subsubsection}{0}

%%%%%%%%%%%%%%%%%%%%%%%%%%%%%%%%%%%%%%%%%%%
%%%%%%%%%%%%%%%%%%%%%%%%%%%%
\singleandthirdspaced

%\begin{document}

\section{Introduction}

Underlying any discussion of ``Beyond the Standard Model Physics" is
the question:  what are the scales of new physics?  For most
particle physicists, theorists and experimentalists, the successes
of the standard model which we have heard about again at this
meeting are impressive, but they are also a source of
enormous frustration.  We believe that the standard model cannot
be complete.  A complete theory, we presume, would not have so
many free parameters; it would incorporate gravity in a consistent
way; it would not be governed by mysterious fine tunings.  To
understand these questions, we have explored grand unification,
supersymmetry, and string theory.  We have contemplated neutrino
masses and compositeness.  As we have heard so beautifully
described at this meeting, we now have good evidence for neutrino
masses.  This certainly means that there is something beyond the
standard model.  What type of new physics is responsible for these
masses -- and what are the associated scales -- are by no means
clear. We have also heard a great deal of discussion of
supersymmetry.  The evidence, here, is tenuous at best, but
only in the case of supersymmetry breaking do we have any
real argument for what the relevant scale should be.  And even
this argument, while oft-repeated, is open to question.   In
string theory, the conventional wisdom has long been that the
fundamental scale of the theory is close to the Planck scale.  But
this has been called into question in recent years, as has been
the significance of the fundamental scale itself. In this talk, I
will not answer the question of what these various
scales may be, but I will try to phrase the
issues as sharply as possible, and discuss some of the suggestions
which have been offered.

Let me turn first to supersymmetry.  Supersymmetry is a beautiful,
hypothetical symmetry of nature.  It has been the subject of
intense theoretical discussion and serious experimental
investigation.  What experimental evidence there is lies in the fact that
the gauge couplings unify at a high energy if one assumes
supersymmetry breaking at about $1~ {\rm TeV}$, and not otherwise.
This is impressive, but hardly compelling by itself.  The real
arguments that nature is supersymmetric, and that supersymmetry is broken at
scales not too different than the weak scale, are theoretical. The
standard model suffers from a difficulty often referred to as the
problem of quadratic divergences or the hierarchy problem.
More simply, it is a collosal failure of dimensional analysis.
In the standard model, the masses of all of the elementary
particles are tied to the mass of the Higgs particle.  Dimensional
analysis suggests that this mass should be of order the largest
scale we know in nature, the Planck scale, and this is manifestly
false.  On the other hand, a
similar argument would suggest that the electron mass should be of
order the $W$ boson mass.  While we don't really understand why
the electron mass is small, we do know that the standard model is
more symmetric if the electron mass is zero, so it is {\it
natural} that the electron mass is light.  More generally, we
believe that the smallness of all of the quark and lepton masses
(except $t$) is due to approximate flavor symmetries. On the other
hand, the standard model {\it does not}
become more symmetric if the Higgs mass is small.
Indeed, if one tries to write a theory with a small
value of the Higgs mass, it will generally receive large radiative
corrections.  If the Higgs particle is to be light as a result
of symmetries, it is necessary to enlarge
the standard model, so that it is supersymmetric.
Obviously, if the laws of nature are supersymmetric, supersymmetry
must be a broken symmetry. If the scale of supersymmetry breaking is not
much different than the scale of weak interactions, than the Higgs
mass is naturally of this order. The only other persuasive
explanation which has been offered for the lightness of the
Higgs particle is technicolor,
the possibility that electroweak symmetry breakdown is due to some
new strong interactions among a new set of fermions know as
techniquarks. Technicolor has fallen into some disfavor in recent
years, and was not discussed at any length at this conference.
Technicolor theories generically have problems with precision
electroweak tests.  It is difficult to construct models which are
compatible with constraints from rare processes. As a result,
there is no ``standard model" of technicolor,
and one can only guess at what the
generic predictions are.  Still, many of us wonder, especially as
LEP and the Tevatron put stronger and stronger limits on
supersymmetric theories, whether some sort of dynamical symmetry
breakdown might be the origin of electroweak symmetry breaking.

Some progress has been made in recent years on a slightly
different form of dynamical symmetry breaking than conventional
technicolor, known as ``top color."  These models also
may require some exotic new physics.  It is interesting that at
this meeting we heard a version of this idea which involves large
compact dimensions to generate the requisite four fermi
operators\cite{dobrescu}.

\section{Supersymmetry}

Their are a number of reasons why so many theorists seem convinced
that supersymmetry will be discovered in the not too distant
future.
\begin{itemize}
\item  Supersymmetry explains
the failure of dimensional analysis to account for
the value of the Higgs mass.
\item  Low energy supersymmetry yields
unification of couplings, without ad hoc
assumptions.
\item  Supersymmetry naturally provides candidates for the dark
matter, without fiddling with parameters.
\item  Supersymmetry fits elegantly into string theory.  If string (M-) theory
is a unique, ultimate theory of gravity and gauge interactions, it
follows that supersymmetry is an essential part of this final
theory.  Low energy supersymmetry emerges naturally from this
story.
\item  Unlike technicolor, complete models exist, whose
phenomenology can be investigated in great detail.
\item  Finally, supersymmetry is a beautiful possible symmetry of
nature; it would be disappointing if nature didn't exploit it.
\end{itemize}

Only the first three arguments suggest that supersymmetry should be
broken at low energies. The last argument, even if persuasive,
does not say that the scale of supersymmetry breaking should be
such that we could hope to observe this symmetry. Clearly some of
these arguments are more compelling than others. The fact that we
can build models easily is hardly, in itself, an argument but it
certainly accounts for some of the subject's appeal. Indeed, all
one needs to do is specify certain ``soft breaking parameters"
(the masses of the superpartners of ordinary fields, as well as
certain trilinear scalar couplings), the particle content (e.g.
the MSSM), and certain discrete symmetries, to completely specify
the phenomenology.

Assuming that nature is supersymmetric at low energies, the most
crucial question in the subject will be: what is the origin of
supersymmetry breaking. In the MSSM, for example, there are $105$
parameters beyond those of the minimal standard model. These
include the masses and mixings of the squarks, sleptons and
gauginos (including CP violating phases), certain trilinear scalar
couplings, and the parameters which determine the Higgs potential.
Without a {\it theory} of supersymmetry breaking, these quantities
are arbitrary (similar to the CKM parameters). However, there are
significant constraints from experiment.
\begin{itemize}
\item  Flavor-changing neutral currents:  these require some
degree of squark and slepton degeneracy, or alignment of these
mass matrices and the quark and lepton matrices.
\item  Limits from direct searches (more shortly)
\item  Hierarchy:  the scale of supersymmetry breaking should not
be too large (presumably not much greater than a TeV, and perhaps
even less?).   Experiments at LEP and the Tevatron are already
squeezing the allowed parameter space.
\end{itemize}

There are a number of traditional approaches to supersymmetry
breaking.  The most popular has been ``Supergravity" (SUGRA)
breaking. In this approach, one assumes that the squarks, sleptons
and gauginos are nearly degenerate at the high scale.  This has
the virtue of simultaneously solving the flavor problem and of
reducing the number of parameters.  I put ``supergravity" in
quotes, however, because nothing about supergravity (gauged
supersymmetry) by itself enforces this structure. Despite frequent
statements to the contrary, there is no sense in which the
universality of gravitational couplings insures equality of squark
and slepton masses.
In string theory (the only theory we have which is locally
supersymmetric and at the same time where one can do calculations)
this degeneracy does hold approximately in some
circumstances\cite{dilatondominated}, but
it is difficult to understand why the true vacuum of the theory
should have this property.
The only other suggestion to solve
this problem of universality has been to postulate flavor
symmetries\cite{susyflavor}.  Finally, in the traditional SUGRA approach, no
attempt is made to understand the origin of supersymmetry
breaking; the scale of this breaking is simply put in by hand.

More recently, a variety of approaches to supersymmetry breaking
have been developed which attempt to explain the large hierarchy
through supersymmetry-breaking dynamics.  Indeed, we know a great
deal about SUSY dynamics, and can often show that SUSY is
dynamically broken. This can provide an explanation of the
hierarchy, since typically \beq M_{susy} \sim M_{p} e^{-a{2 \pi
\over \alpha_{gut}}}. \eeq There are at least three proposals based on this
possibility:
\begin{itemize}
\item Gauge Mediation:  In these theories, supersymmetry is broken
by some new, strong dynamics.  The breaking is communicated to the
superpartners of ordinary fields by gauge interactions.  Masses of
squarks and sleptons are then just functions of their gauge
quantum numbers, leading to adequate suppression of flavor
changing processes, and, more importantly, to predictions. Another
appealing feature of this mechanism is that models exist.  Still,
the scale of the breaking dynamics is not highly constrained.  The
nicest existing models have large scales\cite{largebreaking}.  The
famed CDF $\gamma \gamma e~e E_{miss}$ event, on the other hand,
suggests a low scale, if this is interpreted as production of a
selectron pair (or something similar)\cite{gmsbreviews,pt}. While
a low scale is quite plausible in this picture, there is only one
explicit model, to my knowledge, with a low enough scale to
explain the CDF event\cite{nelsonstrassler}.
\item  String theory:  here, one often speaks of gluino
condensation as a mechanism for supersymmetry
breaking\cite{condensation}. No complete theory of this kind
exists, however.  The general effect of gaugino condensation is to
give a potential for some of the fields, but in regions where one
can calculate, one does not find a stable minimum.  If there are
stable minima at strong coupling, it is difficult to explore them,
and to understand how problems such as flavor changing processes
are solved.\footnote{One suggestion along these lines is known as
the ``racetrack'' model\cite{racetrack}. This is an attempt to
obtain a stable minimum at weak coupling. General issues connected
with this idea will be discussed in a subsequent publication.}
Recently, another suggestion has been put forward, which the
authors refer to as ``sequestered" supersymmetry breaking.  Here,
the idea is that the fields of the standard model live on a brane
(an object which one can think of as analogous to a domain wall),
while the fields responsible for supersymmetry breaking live on a
second brane, well separated from ``ours."  Under certain
circumstances, this idea can lead to a spectrum similar to that
for gauge mediation.  The crucial new element here is a recently
appreciated anomaly, which gives rise to large than expected
gaugino masses\cite{randallsundrum,murayamaluty}.  While complete
models of this kind do not yet exist, they are the subject of
intense investigation.
\item  Compositeness:  Various authors have used ideas connected
with Seiberg duality to construct models with {\it composite}
quarks and leptons.  Typically, the first two generations are
composite, and the squarks quite massive (greater than a TeV or
so).  Gauginos and the third generation squarks are
lighter\cite{ckn,othercomposite}. Flavor changing neutral currents
are suppressed. These models may well be less fine tuned than
gauge mediated models, and are certainly worth further study and
development.
\end{itemize}

It is probably fair to say that no completely compelling model of
supersymmetry breaking yet exists.  We should be striving to
develop a real Supersymmetric Standard Model, including breaking.
For now, we have various approaches to model building, each with
advantages and drawbacks.

Experiment, meanwhile, is narrowing the allowed parameter space.
Indeed, analyses at all of the major experiments reflect
appreciation for the many possible patterns of soft breaking (its
not just your old MSSM anymore).  In the Beyond the Standard Model
Session at this meeting, we heard several talks from LEP and the
Tevatron closing the susy parameter space, i.e. constraining the
possible scales associated with supersymmetry.
%We heard updated
%limits from the Tevatron and LEP at this meeting.
From LEP,
chargino and neutralino limits are now in the range 65.2 and 28.2
GeV, respectively\cite{isavali} (for ${\rm tan}(\beta)>1$); for
sleptons, the limits range from about $75$ GeV to $84$ GeV, with
$R$ parity conserved, and are not much weaker if $R$-parity is
violated\cite{berggren}; CDF can set chargino limits greater than
$100 \rm GeV$ in much of the parameter space\cite{worcester}, with
similar limits on stops\cite{holck}.  DO sets similar strong
limits on charginos ($m_{\chi^{\pm}} > 150$, $m_{\tilde
q}>232$\cite{hedin}).  A number of analyses have been
performed putting limits on gauge mediated models, assuming that
the next to lightest supersymmetric particle decays quickly to
gravitinos. DO sees no evidence for excess
$\gamma \gamma$ events\cite{hedin}; ALEPH places limits on
neutralinos in such a picture of $90$ GeV; indeed, only a tiny
sliver of parameter space is consistent with an $\tilde e^+ \tilde
e^{-}$ interpretation of the famous CDF event\cite{taylor}.  Other
LEP experiments also place strong limits.  Finally, we heard a
talk ruling out any light gluino window\cite{lath}.

These results are interesting and perhaps worrisome for
supersymmetry enthusiasts.  Given the dimensional analysis or fine
tuning argument, the most natural value for the masses of scalars
are of order the $Z$ mass.  Masses much larger than this suggest
the need for fine tuning.  These problems are most severe for
gauge-mediated models.  In these models, one expects that the
masses of the scalar doublets are of order $\alpha_2 \over
\alpha_1$ times the masses of the charged singlet scalars.  Given
limits of order $90$ GeV on these scalars, this suggests that the
doublets should have masses-squared nearly an order of magnitude
larger than $M_Z^2$.    $b \rightarrow s \gamma$ raises
similar puzzles; it would seem that the Higgs particles themselves
must be surprisingly heavy.  Of course, one should always be open
to other possibilities, but
it is  too early to be pessimistic.  We may have simply been a little
unlucky, and there may be cancellations, or it may be that better
models, such as some of the composite models, may explain these
facts naturally.
In the
``SUGRA" picture, things are also getting tighter,%\cite{tuningmssm}
and many workers feel it is necessary to relax some of the
standard assumptions, e.g. unification of gaugino
masses\cite{kanedpf}.

We can turn these arguments around.  If supersymmetry has
something to do with the solution of the hierarchy problem, then
there is an excellent chance it will be observed at TeV II, and it
will certainly be observed at LHC.  At the recent Tevatron
workshop, for example, it was found that the Tevatron may be able
to rule out almost all of the parameter space of the Minimal
Supersymmetric Standard Model.

It is fun to take an optimistic view, and imagine that five years
from now we are in an era where we are unraveling a new,
fundamental symmetry of nature.  Having discovered several states,
theorists are proposing real, plausible, models of the phenomena,
and predicting the masses of new particles.  Experimentalists are,
with regularity, making new discoveries which cause great
confusion.  Congress is holding hearings demanding to know why NLC
construction has barely started.

\section{Neutrino Masses}

Just in case we were beginning to despair that there are no new
scales in nature, the past year has brought us the news of
neutrino masses.  SuperK has strongly confirmed the atmospheric
neutrino deficit, which suggests, along with the solar neutrino
data, that \beq \Delta m_{\mu \tau}^2 \approx (10^{-2}-10^{-3})
{\rm eV}^2 ~~~~~~~~sin^2(2 \theta)= 0.82-1.0. \eeq

These results are beautifully summarized in Gary Feldman's talk at
this meeting.  Here I would just remark on the question:  what do
we learn from this data?  What does this tell us about new scales
in physics?  Neutrino mass is readily accommodated in the standard
model, but it necessarily implies the existence of a new
scale.\footnote{Unless the masses are Dirac masses.  Some
physicists have suggested that we define the standard model to
include right handed neutrinos, with very tiny Yukawa couplings.}
In the standard model, a neutrino mass arises from an operator of
the form \beq {\gamma_{f f^{\prime}} \over M} H L_f H
L_f^{\prime}.\eeq  This is a non-renormalizable operator. It is
technically ``irrelevant," which is another way of saying that the
neutrino mass is very small.  It is often described as arising
from the ``seesaw mechanism," but the appearance of such an
operator is more general, and would be expected to occur in any
theory in which lepton number was violated at some scale $M$.  At
low energies, all we measure is the neutrino mass matrix, \beq
m_{f f^{\prime}} = {\gamma_{f f^{\prime}} v^2 \over M}\eeq

$M$ represents some scale of new physics.  In grand unified
theories, for example, it is often of order the grand unification
scale, or somewhat smaller.   In the seesaw mechanism $M$ is the
mass of a right-handed neutrino, and $\gamma_{\tau \tau}$ might be
of order the $\tau$ Yukawa coupling, squared. Through the years,
numerous papers have been written on such models, starting
with\cite{grs}.  It is relatively easy to cook up models which
explain the data, though the large
mixing angle is surprising.  The real question is whether the model makes
other testable predictions. In the parallel sessions at this
meeting, we heard a description of a model\cite{bwp} which
illustrates the sorts of predictions which may be possible.  This model
naturally gives for $\gamma$ a value of order $m_t^2/v^2$, which
is several orders of magnitude larger than one might have
expected.  A further $M_{GUT}/M_p$ suppression than accounts for
the SuperK observations.  The model also gives a nice picture of
quark and lepton masses and mixings.  It makes some definite
predictions for proton decay: \beq \Gamma^{-1}(p \rightarrow \bar
\nu K^+) \le 10^{33} {\rm yrs} ~~~~~~~ \Gamma^{-1}(p \rightarrow
\bar \mu^+ K^o) \le 10^{34} {\rm yrs} \eeq The SuperK limits on
these modes described at this meeting are $6.8 \times 10^{32}$ and
$4.0 \times 10^{32}$, respectively (in each case nearly an order
of magnitude improvement over the preexisting limit)\cite{superk}.

\section{String (``M") Theory}

String theory has scored many spectacular successes over the past
few years.  Perhaps most dramatically, it appears that the various
known string theories -- the only known consistent theories of
quantum gravity -- are all one and the same theory\cite{wittenusc}.
It is perhaps
not too outlandish to speculate that this structure is the unique
theory of gravity, and the unique possibility for a truly unified
theory. While a complete non-perturbative formulation of the
theory still eludes us, much has been understood about the
fundamental dynamics of this theory, and plausible candidates for
a complete formulation, at least under certain circumstances,
exist. Finally, one of the most serious challenges to reconciling
quantum mechanics and general relativity, the paradox proposed many
years ago by Hawking\cite{hawking} has been at least partially
resolved\cite{sv}.

Still, it has proven difficult to extract quantitative -- or even
convincing qualitative -- predictions from the theory.  One would
like to know, for example:
\begin{itemize}
\item  Does the theory predict low energy supersymmetry?
\item  If so, what is the pattern of soft breakings?
\item  Can one understand the origin of the flavor structure we
observe?
\item  In light of the recent neutrino results, could one account
for the scale of neutrino masses and the mixing pattern?
\end{itemize}

The third and fourth questions are almost certainly premature.
Even the first, often casually cited as a prediction of
superstring theory, is hard to pin down.  The difficulty is not
hard to understand.  It is closely related to the often repeated
statement that string theory is a theory without parameters.  What
is meant by this statement?  What determines the couplings and the
scales of the theory?

If string theory describes nature, all of these quantities must be
determined dynamically.  Within our current understanding the
various coupling constants are determined by the expectation
values of scalar fields, called ``moduli."  One would like to
compute, say, a potential for the moduli, and show that it has a
minimum consistent with the known values of the gauge and Yukawa
couplings.  Naively, on the other hand, one would expect that any
minimum of the potential would occur when all of the couplings
were of order one and the scales comparable.  This can be made
precise.  For couplings, any potential one computes will go to
zero as the couplings go to zero.  Any minimum of the potential
must lie in a regime where the couplings are of order one, i.e.
where the theory is strongly coupled\cite{dineseiberg}. Similarly,
in any ground state with approximate supersymmetry, the potential
must vanish as the compactification radii tend to infinity.

Traditionally, the argument that the compactification scales are
of order one has been made slightly differently.  In the heterotic
string, which was longed viewed as the most promising string
theory phenomenologically,  it was  argued that the Planck scale,
GUT scale, and string scale ($M_s$) must be comparable: $M_p \approx
M_{GUT} \approx M_s$, since \beq \alpha^{-1} = g_s^{-2} V_{comp}
M_s^6. \eeq  Here $g_s$ is the dimensionless string coupling,
$\alpha^{-1} \sim 25$, i.e. a typical grand
unified coupling constant and $V_{comp}$ is the compactification
volume (the compactification radius is essentially the grand
unified scale). If the theory is to be weakly coupled, so that a
perturbative string picture is appropriate, $g_s \le 1$, so the
compactification scale and string tension cannot be very
different.  Newton's constant, similarly, is given by \beq
G_N^{-1} = M_p^2 = V_{comp} M_s^8, \eeq so the Planck mass must be
of order the string scale.

This particular argument is suspect.  After all, it is not obvious
that the string coupling should be small. In light of recent
understandings of duality, one might try to relax this. For
example, the conventional supersymmetric calculation of the grand
unification scale yields a value three orders of magnitude below
the Planck mass. According to our formula for the gauge coupling,
this would mean that the dimensionless string coupling is
enormous. Witten argued, based on this, that a more suitable
starting point for string
phenomenology would be the strong coupling limit of the
heterotic string\cite{wittencompact}. Horava and Witten had
previously shown that at strong coupling, the heterotic string
appears eleven dimensional, with two ten dimensional ``walls" at
the end of the world\cite{hv}.  If one takes literally these formulas, one
finds that the eleven dimensional Planck mass, the GUT scale, and
the compactification radius are all of order $10^{16}$ GeV.  The
size of the eleventh dimension, $R_{11}$, is of order $70$ times
the GUT scale.  In this picture, the four dimensional Planck scale
is not particularly fundamental. It is of order $R^{3/4}$ times
the fundamental eleven dimensional scale.

In this picture, no parameter is particularly large.  Still, it is
puzzling that a pure number of order $70$ should arise.  Indeed,
as $R_{11}$ grows, the theory effectively becomes five dimensional
(if the other compactification lengths are held fixed).
Supersymmetry in five dimensions essentially forbids a potential
for $R_{11}$, i.e. the potential must vanish in this limit, and it
is hard to understand why stabilization does not occur for $R_{11}
\sim 1$\cite{banksdinescales}. This is similar to the puzzle in
the string theory picture of understanding why, in a strongly
coupled theory, the gauge couplings should be small and
unified\cite{coping}.  Still, the picture is appealing and
suggestive.  Given the vast number of string compactifications,
one can certainly imagine that some involve large pure numbers.
Perhaps anthropic or other considerations might explain why one
with a small coupling is favored.  We are not currently in a
position to pose these questions precisely.

\section{A More Radical Proposal:  Strings at the TeV Scale}

In the last section, we discussed Witten's suggestion that the
fundamental scale of physics lies at a scale well below the four
dimensional Planck scale.  This idea has been taken much further
in \cite{lykken,dimopoulos,precursors,otherextradim}.  These
authors explore the possibility that the string scale is of order
a TeV. This has the virtue that, because the fundamental scale is
of order the scale of electroweak symmetry breaking, one avoids
the failure of dimensional analysis (hierarchy problem)
which we discussed earlier.  In this picture, the standard model
fields live, again, on a brane or domain wall. As a result, the
gauge couplings are not particularly sensitive to the size of the
extra dimensions.

The first interesting question is the value of the other scales.
In this picture, the four dimensional Planck scale is again a
derived quantity.  It is given by a formula of the form \beq
M_p^2 = V_D M_s^8 \eeq in a ten dimensional picture; here $V_D$ is
the volume of the internal space (in an eleven dimensional
picture, the relevant scale is the eleven dimensional Planck
scale).   If $n$ dimensions are large, while the others
are comparable to the fundamental scale (now assumed
to be a TeV), then $V_D \sim r^n$ in $TeV$ units, and
\beq
r \approx (M_p/TeV)^{2/n} (TeV)^{-1}.
\eeq
If the six compact dimensions are of comparable size,
than the compactification scale is large, $r \sim ({\rm
(Mev})^{-1}$.  If only two of the dimensions are large, than $r
\sim {\rm mm}$! This is  quite a spectacular result.   It means,
for example, that if one can perform measurements at scales
slightly less than a millimeter, one should see a change in
Newton's law of gravitation from $1/r^2$ to that appropriate to
six dimensions, $1/r^4$!

For general $n$, the assumption of a low string scale makes other
dramatic predictions.  For example, individual Kaluza-Klein states
couple with gravitational strength, but there are lots of them. At
energy scales above the compactification scale, the cross section
is that appropriate to a theory of $4+n$ dimensions.  It grows
rapidly with energy.  One can, in fact, set interesting limits on
these theories from various processes involving large amounts of
missing energy\cite{accelerators}.

Apart from direct search experiments, there are many other
constraints.  The most obvious is baryon number violation. Baryon
number violation is suppressed only by powers of $M\sim TeV$, so
it is necessary to suppress many operators.  In
conventional supersymmetric
models, it is also necessary to suppress certain operators of dimension
four and five; this
is usually achieved with discrete symmetries.  In the
case of a TeV fundamental scale, one needs to suppress operators
up to something like dimension 12, so more powerful symmetries are
needed.  Still, one can postulate discrete symmetries or other
possibilities which would do the job.  If our goal is to rule out
the possibility of large dimensions,
then, proton decay is not sufficient.
%  One needs
%to understand other flavor problems as well.
Accelerator searches
constrain the scale to be greater than about $2~ {\rm Tev}$.
Flavor-changing neutral currents require some approximate flavor
symmetries\cite{dimopoulosflavor}, and then one still probably
requires a scale greater than about $5-10~ {\rm TeV}$\cite{bdn}.
Astrophysics constraints are particularly severe in the case of a
low number of large compact dimensions. For $n=2$, in particular,
the scale must be greater than about $50$ TeV; otherwise, emission
of light particles by SN1987a is too efficient\cite{supernova}.

From the work which has been done in this subject, then, it is
clear that one cannot so easily rule out the possibility of a low
string scale, and surprisingly large dimensions. The arguments
described above for the case $n=2$
suggest that it is unlikely one will observe extra
dimensions through a change in the power in Newton's
law in Cavendish experiments. It is still possible that one will
see alterations in gravity due to light particles\cite{bdn}.  One
is unlikely, in this case, to be able to observe dramatic effects in
accelerators.

It is hard to imagine a more exciting discovery
than large
extra dimensions, and it is clearly worthwhile to
search for phenomena which might indicate their existence.
But the fact that large dimensions are possible does not necessarily
mean that they are plausible.
In
assessing the likelihood of large dimensions, there are two
questions which seem appropriate:
\begin{itemize}
\item  Is their any puzzle which large dimensions settle.
\item  Are their plausible dynamics which might give rise to such
large dimensions.
\end{itemize}
To the first question, to my knowledge, the only motivation which
has been suggested is the fine-tuning problem.  This requires that
the scale be close to the weak scale.  The fact that for $n=2$ the
scale already has to be so much larger argues strongly against
this possibility.  For $n>2$, the limits on the scale are already
uncomfortably large (suggesting fine tuning of better than a part
in $100$).  These constraints will get stronger over time.   For
the second problem, one typically finds that one must introduce a
large number in a rather ad hoc fashion.  Otherwise, not
surprisingly, one finds that the scales all come out of order one.
In some cases\cite{bkn}, the required coincidence is not much
worse than that required in other ideas about fixing string
moduli.  Various other coincidences are also required.  The
cosmology of these theories is also problematic\cite{largedcosmo}.
My personal view, based on these observations, is that large
dimensions are not terribly likely, but that one should keep an
open mind, both theoretically and experimentally.

It is worth mentioning variants on the brane ideas, which are not
quite as extreme, but which raise similar theoretical
issues\cite{randallsundrum,ibanezetal}.

\section{Flavor Physics}

It should be stressed that most ideas about physics beyond the
standard model, at scales of order the weak scale, have
implications for flavor physics.  Many proposals for low energy
realizations of supersymmetry predict rare processes should
be near the current limits.
Similarly, the large dimension ideas which
we discussed above almost inevitably to predict
rates for rare $K$ decays, $D \bar D$ mixing,
and lepton number violation at levels not too far from
the present limits.  If the scale is low, the large
dimension picture tends to
predict that the CKM matrix is real\cite{bdn}, so that observed CP
violation should occur through higher dimension operators.  Rare K
processes should be close to the experimental limits.  Lepton
violating processes might be forbidden by discrete symmetries
(though this may be hard to reconcile with the observation of
neutrino mass), but otherwise should be near the limits. As
stressed in Marciano's talk\cite{marciano}, effects may even be
observable in the muon $g-2$.

\section{Maximally Enhanced Symmetries}

Duality is a very attractive idea.  In some cases, dualities
represent actual symmetries.  A beautiful example is suggested by
the discussion of electric-magnetic duality in classical
electrodynamics, which you can find in Jackson's chapter 6.  Under
this symmetry, which everyone has pondered at one time or another,
one has \beq \vec E \rightarrow \vec B ~~~~~~\vec B \rightarrow -
\vec E ~~~~ {e^2 \rightarrow {2 \pi \over e^2}}. \eeq When Jackson
wrote his text, one could only speculate on the possibility of
such a symmetry.  Now we know that this is a true symmetry of
string theory and some field theories.  Apart from being
theoretically satisfying, this observation raises the question:
could it be that the state we observe in nature is a {\it fixed
point} of such a symmetry\cite{dns}.  Such a hypothesis is
plausible for at least two reasons:
\begin{itemize}
\item  It is technically natural; such enhanced symmetry points
are necessarily stationary points of the effective action of the
theory.
\item  It solves one of the most serious problems of any string
cosmology, the ``moduli problem."
\end{itemize}
Such a hypothesis is predictive:  it suggests supersymmetry, with
supersymmetry breaking at low energies, presumably through something
like gauge mediation.  All of the couplings are
of order one; the scales are comparable.  One disturbing feature
of this hypothesis is that generically one has $\alpha \sim 1$ at
such points.  It is not known whether one can find points with
small couplings.

\section{A Parting, Cautionary Note}

The hierarchy problem underlies much of our thinking about physics
beyond the standard model.  It is the only argument that new
physics should appear at some particular, accessible energy scale.
However, there is another stupendous failure of dimensional
analysis, for which we have no persuasive explanation.
The problem is the
cosmological constant problem.  Why is $\Lambda=0$, or perhaps
$\Lambda \approx 10^{-47}GeV^4$, as suggested by the recent
supernova observations?
This
suggests that there might be some other explanation of hierarchies
which we have simply not thought of.

Perhaps the most interesting recent proposal in this regard is
that of \cite{ks}.  These authors have constructed models without
supersymmetry which have $\Lambda=0$, at least in low orders of
perturbation theory (I should note that, as of yet, these are toy
models, which are certainly not realistic).  There are heuristic
arguments that these statements are true to all orders of
perturbation theory.  Harvey has argued that at least in some
cases, duality suggests that there can be exponentially small
cosmological constants\cite{harvey}.  Does one
know what might be the low energy consequences of such a picture?
Certainly not yet, but -- be alert for surprises.

\noindent {\bf Acknowledgements:}

\noindent

I am grateful to Nima Arkani-Hamed, Howard Haber, Hitoshi Murayama,
Ann Nelson and Lisa Randall for discussions during the
preparation of this talk, and to Ann Nelson for a careful
reading the manuscript (I am responsible for any errors
and omissions). This
work was supported in part by the U.S. Department of Energy.

%%%%%%%%%%%%%%%%%%%%%%%%%%%%%%
%  Bibliography
%%%%%%%%%%%%%%%%%%%%%%%%%%%%%%

\end{document}